# Relativistic quark model and pentaquark spectroscopy.

Gerasyuta S.M. [1,2,3], Kochkin V.I. [1]

1. Department of Theoretical Physics, St. Petersburg State University, 198904, St. Petersburg, Russia.
2. Department of Physics, LTA, 194021 St. Petersburg, Russia.
3. Forschungzentrum Julich, Institut fur Kernphysik (Theorie) D-52425 Julich, Germany.

Abstract.

The relativistic five-quark equations are found in the framework of the dispersion relation technique. The solutions of these equations using the method based on the extraction of leading singularities of the amplitudes are obtained. The five-quark amplitudes for the low-lying pentaquarks are calculated under the condition that flavor $SU(3)$ symmetry holds. The poles of five-quark amplitudes determine the masses of the lowest pentaquarks. The mass spectra of pentaquarks which contain only light quarks are calculated. The calculation of pentaquark amplitudes estimates the contributions of three subamplitudes: molecular subamplitude $BM$, $Mqqq$ subamplitude and $Bq\bar{q}$ subamplitude. The main contributions to the pentaquark amplitude are determined by the subamplitudes, which include the meson states M.

PACS numbers: 11.55 Fv, 12.39 Ki, 12.39 Mk, 12.40 Yx

I. Introduction

The existence of particles made of more than three quarks is an important issue of QCD inspired models. The role of stable exotic against strong decays is crucial for understanding of some aspects of the strong interactions. Their properties could be an important test for the validity of various quark models. The five-quark systems (pentaquarks) was proposed independently Gignoux, Silvestre-Brac, Richard [1] and Lipkin [2]. The more realistic calculations [3] which take into account the flavor $SU(3)$ - breaking, lead to stable systems against strong decays. In the De Rujula-Georgi-Glashow model [4] the stable pentaquarks have negative parity and require strangeness [3].

Accoding to recent experiments, no convincing evidence for the production of charmed-strange pentaquarks has been observed either. However the existence of pentaquarks is not ruled out. The analysis done so far can provide a good starting point for the future search in high statistics charm experiments at CERN [5] or Fermilab [6]. The extension of the Skyrmion approach to the heavy flavor sectors [7, 8] allowed to calculate the spectra of the low-lying pentaquarks containing charm and bottom antiquarks. The conclusion of this model are similar to those of chiral model with Goldstone boson exchange (GBE) [9]. In the GBE the candidates of stability are not necessarely strange and have positive parity [10]. The Skyrmion approach allowed to calculate the low-lying exotic baryon $Z^+$ (spin ½, isospin 0, strangeness +1) with mass 1530 MeV [11]. The binding of pentaquarks due to the long range one pion exchange has been observed to show a molecular type structure [12]. The lattice gauge calculations became also recently avaible [13] and they may shed additional light on the interaction of quarks.

In [14, 15] a relativistic generalization of the three-body Faddeev - Yakubovsky equations was obtained in the form of dispersion relations in the pair energy of two interacting particles. The mass spectrum of S-wave baryons including u, d and s quarks was calculated by method based on isolating of leading singularities in the amplitude. We found the approximate solution of integral three-quark equations by taking into account two-particle and triangle singularities, all the weaker ones being neglected. If we considered the approximation, which corresponds to taking into account two-body and triangle singularities, and defined all the smooth functions of the subenergy variables (as compared with the singular part of the amplitude) in the middle point of physical region of Dalitz-plot, then the problem reduces to one of solving simple algebraic system equations.

In our paper [16] the relativistic generalization of the four-body Faddeev - Yakubovsky type equations are represented in the form of dispersion relations of the two-body subenergy. We investigated the relativistic scattering four-body amplitudes of the constituent quarks of two flavors (u, d). The poles of these amplitudes determine the masses of the lowest hybrid mesons. The constituent quark is color triplet and quark amplitudes obey the global color symmetry. We used the results of the bootstrap quark model [17] and introduced the $q\bar{q}$ – state in the color octet channel with $J^{PC} = 1^{--}$ and isospin I = 0. This bound state is identified as a constituent gluon. In our model we take into account the color state with $J^{PC} = 1^{--}$ and isospin I = 1, which determines with the constituent gluon the hybrid state. In addition, $q\bar{q}q\bar{q}$ states are also predicted.

We derived the mixing of the hybrid and $q\bar{q}q\bar{q}$ states. This state was called the hybrid meson. The mass spectrum of lowest hybrid mesons with isospin I = 1 both with exotic quantum numbers (non -$q\bar{q}$) $J^{PC} = 1^{-+}$, $0^{--}$ and ordinary quantum numbers $J^{PC} = 0^{++}$, $1^{++}$, $2^{++}$, $0^{-+}$, $1^{--}$ was calculated. The important result of this model is the calculation of hybrid meson amplitudes, which contain the contribution of two subamplitudes: four-quark amplitude and hybrid amplitude. The main contribution corresponds to the four-quark amplitude. The hybrid amplitude gives rise to only less 40 % of the hybrid meson contribution.

In our paper [18] the relativistic generalization of five-quark equations (like Faddeev – Yakubovsky approach) are constructed in the form of the dispersion relation. The five-quark amplitudes for the low-lying hybrid baryons contain only light quarks and are calculated under the condition that flavor SU(3) symmetry holds. In should be noted, that the calculated masses of low-lying hybrid baryons agree with data [19] and with the results obtained in the flux-tube model [20].

In our relativistic quark model with four-fermion interaction the octet color $q\bar{q}$ bound state was found, which corresponds to the constituent gluon G with mass $M_G$ = 0.67 GeV [17]. This approach is similar to the large $N_c$ limit [21-23]. In diquark channel we have the diquark level D with $J^P = 0^+$ and the mass $m_{ud}$ = 0.72 GeV (in the color state $\bar{3}_c$). The diquark state with $J^P = 1^+$ in color state $\bar{3}_c$ also has an attractive interaction, but smaller than that of the diquark with $J^P = 0^+$, therefore there is only the correlation of quarks, not a bound state [17].

The calculated five-quark amplitude consists of four subamplitudes: $qDG$, $qqqG$, $D\bar{q}D$ and $qq\bar{q}D$, where $D$ and $G$ are the diquark state and exited constituent gluon state respectively. The main contributions to the hybrid baryon amplitude are determined by the subamplitudes, which include the exited gluon states.

The present paper is devoted to the construction of relativistic five-quark equations for the pentaquarks. The five-quark amplitudes for the lowest pentaquarks contain only light quarks and take into account the flavor SU(3) symmetry. The poles of these amplitudes determine the masses of the low-lying pentaquarks. The constituent quark is the color triplet and the quark amplitudes obey the global color symmetry. The interesting result of this model is the calculation of pentaquark amplitudes which contain the contribution of three subamplitudes: molecular subamplitude $BM$, $Mqqq$ subamplitude and $Bq\bar{q}$ subamplitude. Here $B$ corresponds to the lowest baryon (nucleon and $\Delta$–isobar baryon). $M$ are the low-lying mesons with the quantum numbers: $J^{PC} = 0^{++}, 1^{++}, 2^{++}, 0^{-+}, 1^{--}$ and isospin I = 0. We call the pentaquark with $J^P = \frac{1}{2}^+$ as the $N$ pentaquark and the pentaquark with $J^P = \frac{3}{2}^+$ as the $\Delta$–isobar pentaquark.

The mass values of the low-lying pentaquarks are calculated (Table 1, 2). The lowest mass of $N$ pentaquark with $J^P = \frac{1}{2}^+$ is equal M=1553 MeV. The pentaquark amplitudes take into account the contribution of three subamplitudes. The main contributions to the pentaquark amplitude are determined by subamplitudes which include the low-lying meson with $J^{PC} = 0^{++}, 1^{++}, 2^{++}, 0^{-+}, 1^{--}$.

The paper is organized as follows.

After this introduction, we discuss the five-quark amplitudes which contain only light quarks (section 2).

In the section 3, we report our numerical results (Tables 1, 2) and the last section is devoted to our discussion and conclusion.

In the Appendix A we give the relations, which allow to pass from the integration of the cosines of the angles to the integration of the subenergies.

In the Appendix B we describe the integration contours of functions $J_1$, $J_2$, $J_3$, which are determined by the interaction of the five quarks.

In the Appendix C we obtain the determinant of the algebraic equations, which allows one to calculate the mass spectra of the pentaquarks.

In the Appendix D the quark-quark and quark-antiquark vertex functions and phase spaces for the pentaquarks are given respectively (Tables 3, 4).

## II. Pentaquark amplitudes.

We derived the relativistic five-quark equations in the framework of the dispersion relation technique. For the sake of simplicity one considers the case of the $SU(3)_f$ - symmetry, that the masses of all quarks are equal. We use only planar diagrams, the other diagrams due to the rules of $1/N_c$ expansion [21-23] are neglected. The correct equations for the amplitude are obtained by taking into account the all possible subamplitudes. It corresponds to the division complete system into subsystems from the smaller number of particles. Then one should represent five-particle amplitude as a sum of ten subamplitudes: $A = A_{12} + A_{13} + A_{14} + A_{15} + A_{23} + A_{24} + A_{25} + A_{34} + A_{35} + A_{45}$. In our case all particles are identical, therefore we need to consider only one group of diagrams and the amplitude corresponding to them, for example $A_{12}$. The set of diagrams associated with the amplitude $A_{12}$ can be further broken down into three groups corresponding to amplitudes $A_1(s, s_{1234}, s_{12}, s_{34})$, $A_2(s, s_{1234}, s_{12}, s_{123})$, $A_3(s, s_{1234}, s_{25}, s_{125})$ (Fig. 1). The antiquark is shown by the arrow, the other lines correspond to the quarks. The coefficients are determined by the permutation of quarks [24, 25].

In order to represent the subamplitudes $A_1(s, s_{1234}, s_{12}, s_{34})$, $A_2(s, s_{1234}, s_{12}, s_{123})$, and $A_3(s, s_{1234}, s_{25}, s_{125})$ in form of the dispersion relation it is necessary to define the amplitudes of quark-quark and quark-antiquark interaction $b_n(s_{ik})$. The quark amplitudes $q\bar{q} \to q\bar{q}$ and $qq \to qq$ are calculated in the framework of the dispersion N/D method with the input four-fermion interaction with quantum numbers of the gluon [17]. We use the results of our relativistic quark model [17] and write down the pair quarks amplitude in the form:

$$b_n(s_{ik}) = \frac{G_n^2(s_{ik})}{1 - B_n(s_{ik})}, \tag{1}$$

$$B_n(s_{ik}) = \int_{4m^2}^{\Lambda} \frac{ds'_{ik}}{\pi} \frac{\rho_n(s'_{ik}) G_n^2(s'_{ik})}{s'_{ik} - s_{ik}}. \tag{2}$$

Here $s_{ik}$ is the two-particle subenergy squared, $s_{ijk}$ corresponds to the energy squared of particles $i$, $j$, $k$, $s_{ijkl}$ is the four-particle subenergy squared and $s$ is the system total energy squared. $G_n(s_{ik})$ are the quark-quark and quark-antiquark vertex functions (Table 3). $B_n(s_{ik})$, $\rho_n(s_{ik})$ are the Chew-Mandelstam functions with the cut – off $\Lambda$ [26] and the phase spaces respectively (Appendix D, Table 4). There n=1 corresponds to $qq$-pair with $J^P = 0^+$ in the $\bar{3}_c$ color state, n=2 describes $qq$-pair with $J^P = 1^+$ in the $\bar{3}_c$ color state and n=3 defines the $q\bar{q}$-pairs, with correspond to the mesons with quantum numbers: $J^{PC} = 0^{++}, 1^{++}, 2^{++}, 0^{-+}, 1^{--}$ and isospin I = 0.

In the case in question the interacting quarks do not produce a bound state, therefore the integration in (3) - (5) is carried out from the threshold $4m^2$ to the cut-off $\Lambda$. The integral equation systems, corresponding to Fig. 1 (the meson state with $J^{PC} = 0^{++}$ and diquark with $J^P = 0^+$), can be described as:

$$A_1(s, s_{1234}, s_{12}, s_{34}) = \frac{\lambda_1 B_3(s_{12}) B_1(s_{34})}{[1 - B_3(s_{12})][1 - B_1(s_{34})]} + 6\hat{J}_2(3,1) A_3(s, s_{1234}, s'_{23}, s'_{234}) + \\ + 2\hat{J}_2(3,1) A_2(s, s_{1234}, s'_{13}, s'_{134}) + 6\hat{J}_1(3) A_2(s, s_{1234}, s'_{15}, s_{125}) + 2\hat{J}_1(3) A_3(s, s_{1234}, s'_{25}, s_{125}) \tag{3}$$

$$A_2(s, s_{1234}, s_{12}, s_{123}) = \frac{\lambda_2 B_3(s_{12})}{1 - B_3(s_{12})} + 4\hat{J}_3(3) A_1(s, s_{1234}, s'_{13}, s'_{24}), \tag{4}$$

$$A_3(s,s_{1234},s_{25},s_{125}) = \frac{\lambda_3 B_1(s_{25})}{1-B_1(s_{25})} + 2\hat{J}_3(1)A_1(s,s_{1234},s'_{35},s'_{21}), \quad (5)$$

were $\lambda_i$ are the current constants. We introduced the integral operators:

$$\hat{J}_1(l) = \frac{G_l(s_{12})}{[1-B_l(s_{12})]} \int_{4m^2}^{\Lambda} \frac{ds'_{12}}{\pi} \frac{G_l(s'_{12})\rho_l(s'_{12})}{s'_{12}-s_{12}} \int_{-1}^{+1} \frac{dz_1}{2}, \quad (6)$$

$$\hat{J}_2(l,p) = \frac{G_l(s_{12})G_p(s_{34})}{[1-B_l(s_{12})][1-B_p(s_{34})]} \times$$
$$\times \int_{4m^2}^{\Lambda} \frac{ds'_{12}}{\pi} \frac{G_l(s'_{12})\rho_l(s'_{12})}{s'_{12}-s_{12}} \int_{4m^2}^{\Lambda} \frac{ds'_{34}}{\pi} \frac{G_p(s'_{34})\rho_p(s'_{34})}{s'_{34}-s_{34}} \int_{-1}^{+1} \frac{dz_3}{2} \int_{-1}^{+1} \frac{dz_4}{2}, \quad (7)$$

$$\hat{J}_3(l) = \frac{G_l(s_{12},\widetilde{\Lambda})}{1-B_l(s_{12},\widetilde{\Lambda})} \times$$
$$\times \frac{1}{4\pi} \int_{4m^2}^{\widetilde{\Lambda}} \frac{ds'_{12}}{\pi} \frac{G_l(s'_{12},\widetilde{\Lambda})\rho_l(s'_{12})}{s'_{12}-s_{12}} \int_{-1}^{+1} \frac{dz_1}{2} \int_{-1}^{+1} dz \int_{z_2^-}^{z_2^+} dz_2 \frac{1}{\sqrt{1-z^2-z_1^2-z_2^2+2zz_1z_2}}, \quad (8)$$

were $l,p$ are equal 1 or 3. If we use the diquark state with $J^P = 1^+$ and the meson with $J^{PC} = 0^{++}, 1^{++}, 2^{++}, 0^{-+}, 1^{--}$, $l,p$ are equal 2 or 3. There $m$ is a quark mass.

Hereafter we suggest that some unknown (not large) contribution from small distances which might be taken into account by the cut-off procedure. In the (6) – (8) we choose the "hard" cutting, but we can use also the "soft" cutting, for instance $G_n(s_{ik}) = G_n \exp(-(s_{ik}-4m^2)^2/\Lambda^2)$. It does not change essentially the calculated mass spectrum.

In the equations (6) and (8) $z_1$ is the cosine of the angle between the relative momentum of the particles 1 and 2 in the intermediate state and the momentum of the particle 3 in the final state, is taken in the c.m. of particles 1 and 2. In the equation (8) $z$ is the cosine of the angle between the momenta of the particles 3 and 4 in the final state, is taken in the c.m. of particles 1 and 2. $z_2$ is the cosine of the angle between the relative momentum of particles 1 and 2 in the intermediate state and the momentum of the particle 4 in the final state, is taken in the c.m. of particles 1 and 2. In the equation (7): $z_3$ is the cosine of the angle between relative momentum of particles 1 and 2 in the intermediate state and the relative momentum of particles 3 and 4 in the intermediate state, is taken in the c.m. of particles 1 and 2. $z_4$ is the cosine of the angle between the relative momentum of the particles 3 and 4 in the intermediate state and momentum of the particle 1 in the intermediate state, is taken in the c.m. of particles 3, 4.

Using the relation of Appendix A we can pass from the integration of the cosines of the angles to the integration of the subenergies.

Let us extract two-particle singularities in the amplitudes $A_1(s, s_{1234}, s_{12}, s_{34})$, $A_2(s, s_{1234}, s_{12}, s_{123})$ and $A_3(s, s_{1234}, s_{25}, s_{125})$:

$$A_1(s, s_{1234}, s_{12}, s_{34}) = \frac{\alpha_1(s, s_{1234}, s_{12}, s_{34}) B_3(s_{12}) B_1(s_{34})}{[1 - B_3(s_{12})][1 - B_1(s_{34})]} . \tag{9}$$

$$A_2(s, s_{1234}, s_{12}, s_{123}) = \frac{\alpha_2(s, s_{1234}, s_{12}, s_{123}) B_3(s_{12})}{1 - B_3(s_{12})} , \tag{10}$$

$$A_3(s, s_{1234}, s_{25}, s_{125}) = \frac{\alpha_3(s, s_{1234}, s_{25}, s_{125}) B_1(s_{25})}{1 - B_1(s_{25})} , \tag{11}$$

We do not extract three- and four-particle singularities, because they are weaker than two-particle singularities.

We used the classification of singularities, which was proposed in paper [27] for the two and three particle singularities. The construction of approximate solution of the (3) - (5) is based on the extraction of the leading singularities of the amplitudes. The main singularities in $s_{ik} \approx 4m^2$ are from pair rescattering of the particles i and k. First of all there are threshold square-root singularities. Also possible singularities are pole singularities which correspond to the bound states. The diagrams of Fig.1 apart from two-particle singularities have the triangular singularities, the singularities define the interaction of four and five particles. Such classification allowed us to find the corresponding solution of (3) - (5) by taking into account some definite number of leading singularities and neglecting all the weaker ones. We considered the approximation which defines two-particle, triangle, four- and five-particle singularities. The functions $\alpha_1(s, s_{1234}, s_{12}, s_{34})$, $\alpha_2(s, s_{1234}, s_{12}, s_{123})$ and $\alpha_3(s, s_{1234}, s_{25}, s_{125})$ are the smooth functions of $s_{ik}$, $s_{ijk}$, $s_{ijkl}$, $s$ as compared with the singular part of the amplitudes, hence they can be expanded in a series in the singularity point and only the first term of this series should be employed further. Using this classification one define the reduced amplitudes $\alpha_1$, $\alpha_2$, $\alpha_3$ as well as the B-functions in the middle point of the physical region of Dalitz-plot at the point $s_0$:

$$s_0^{ik} = s_0 = \frac{s + 15m^2}{10} \tag{12}$$

$$s_{123} = 3s_0 - 3m^2, \quad s_{1234} = 6s_0 - 8m^2$$

Such a choice of point $s_0$ allows to replace the integral equations (3) - (5) (Fig. 1) by the algebraic equations (13) - (15) respectively:

$$\alpha_1 = \lambda_1 + 6\alpha_3 J_2(3,1,1) + 2\alpha_2 J_2(3,1,3) + 6\alpha_2 J_1(3,3,1) + 2\alpha_3 J_1(3,1,1), \tag{13}$$
$$\alpha_2 = \lambda_2 + 4\alpha_1 J_3(3,3,1), \tag{14}$$
$$\alpha_3 = \lambda_3 + 2\alpha_1 J_3(1,1,3). \tag{15}$$

We use the functions $J_1(l,p)$, $J_2(l,p,r)$, $J_3(l,p,r)$ ($l, p, r = 1, 2, 3$):

$$J_1(l,p) = \frac{G_l^2(s_0^{12})B_p(s_0^{15})}{B_l(s_0^{12})} \int_{4m^2}^{\Lambda} \frac{ds'_{12}}{\pi} \frac{\rho_l(s'_{12})}{s'_{12} - s_0^{12}} \int_{-1}^{+1} \frac{dz_1}{2} \frac{1}{1 - B_p(s'_{15})}, \tag{16}$$

$$J_2(l,p,r) = \frac{G_l^2(s_0^{12})G_p^2(s_0^{34})B_r(s_0^{13})}{B_l(s_0^{12})B_p(s_0^{34})} \times$$
$$\times \int_{4m^2}^{\Lambda} \frac{ds'_{12}}{\pi} \frac{\rho_l(s'_{12})}{s'_{12} - s_0^{12}} \int_{4m^2}^{\Lambda} \frac{ds'_{34}}{\pi} \frac{\rho_p(s'_{34})}{s'_{34} - s_0^{34}} \int_{-1}^{+1} \frac{dz_3}{2} \int_{-1}^{+1} \frac{dz_4}{2} \frac{1}{1 - B_r(s'_{13})} \tag{17}$$

$$J_3(l,p,r) = \frac{G_l^2(s_0^{12}, \tilde{\Lambda})B_p(s_0^{13})B_r(s_0^{24})}{1 - B_l(s_0^{12}, \tilde{\Lambda})} \frac{1 - B_l(s_0^{12})}{B_l(s_0^{12})} \times$$
$$\times \frac{1}{4\pi} \int_{4m^2}^{\tilde{\Lambda}} \frac{ds'_{12}}{\pi} \frac{\rho_l(s'_{12})}{s'_{12} - s_0^{12}} \int_{-1}^{+1} \frac{dz_1}{2} \int_{-1}^{+1} dz \int_{z_2^-}^{z_2^+} dz_2 \frac{1}{\sqrt{1 - z^2 - z_1^2 - z_2^2 + 2zz_1z_2}} \frac{1}{[1 - B_p(s'_{13})][1 - B_r(s'_{24})]} \tag{18}$$

The other choices of point $s_0$ do not change essentially the contributions of $\alpha_1$, $\alpha_2$, and $\alpha_3$, therefore we omit the indexes $s_0^{ik}$. Since the vertex functions depend only slightly on energy it is possible to treat them as constants in our approximation and determine them in a way similar to that used in [28, 29].

The integration contours of functions $J_1$, $J_2$, $J_3$ are given in the Appendix B (Figs. 3, 4, 5). The equations, which are similar to (13) – (15), correspond to other low-lying mesons with isospin I = 0, $J^{PC} = 0^{++}, 1^{++}, 2^{++}, 0^{-+}, 1^{--}$ and diquarks with $J^P = 0^+, 1^+$ (graphic equations Fig.1, 2) are considered in the Appendix C.

The solutions of the system of equations (Appendix C) are considered as:

$$\alpha_i(s) = F_i(s, \lambda_i) / D(s), \tag{19}$$

where zeros of $D(s)$ determinants define the masses of bound states of pentaquarks. $F_i(s, \lambda_i)$ are the functions of $s$ and $\lambda_i$. The functions $F_i(s, \lambda_i)$ determine the contributions of subamplitudes to the pentaquark amplitude.

### III. Calculation results.

The poles of the reduced amplitudes $\alpha_1$, $\alpha_2$, $\alpha_3$, correspond to the bound states and determine the masses of $N$ and $\Delta$ – isobar pentaquarks. In the considered calculation the quark mass are not fixed. In order to fix anyhow $m$, we assume $m = 405\, MeV$ $(m \geq \frac{1}{5} m_{\frac{5}{2}^+}(1990))$. The model has only one new parameter as compared to our model of hybrid baryons [18]. The gluon coupling constant $g = 0.357$ is determined by the fixing of $N$ pentaquark mass $m_{\frac{5}{2}^+}(1990)$. The cut-off parameters are similar to the paper [18]: the cut-off parameters $\Lambda_{0^+} = 22$ and $\Lambda_{1^+} = 32.4$ for nucleon and $\Delta$ – isobar pentaquarks respectively. The calculated mass values of low-lying nucleon and $\Delta$ – isobar pentaquarks are shown in Tables 1, 2. We found the lowest masses of $N$ pentaquarks with $J^P = \frac{1}{2}^-$ M=1378 MeV, $J^P = \frac{1}{2}^+$ M=1553 MeV and $\Delta$ – isobar pentaquarks with $J^P = \frac{3}{2}^-$ M=1150 MeV, $J^P = \frac{3}{2}^+$ M=1290 MeV. If we increase the quark mass, the masses of the lowest $\Delta$ – isobar pentaquarks can be increased, but the masses of the pentaquarks will be most of the calculated masses (Tables 1, 2). The low-lying $\Delta$ – isobar pentaquark masses are smaller than the $N$ pentaquark masses. It is depended on the different interactions in the diquark channels $J^P = 0^+, 1^+$. The calculated values of the pentaquark masses are compared to the experimental data [19]. We predict the degeneracy of some states. The calculation of pentaquark amplitude estimates the contributions of three subamplitudes. The main contributions to the pentaquark amplitude are determined by the subamplitudes, which include the low-lying meson states. The Tables 1, 2 show the contributions of the following subamplitudes: $A_1$ ($BM$), $A_2$ ($Mqqq$), $A_3$ ($Mq\bar{q}$). We sound that the contributions of $A_1$ and $A_2$ subamplitudes are about 40-50 % of the pentaquark contribution. The contribution of the subamplitude $A_3$ is less than 15 % of the pentaquark amplitude. The mass values of $\Delta$ – isobar pentaquarks with $J^P = \frac{3}{2}^-$ M=1150 MeV end $J^P = \frac{3}{2}^+$ M=1290 MeV are depended on the large molecular contributions 71 % and 75 % of the $\Delta$ – isobar pentaquark amplitudes (Table 2). The lightest nucleon pentaquark with $J^P = \frac{1}{2}^+$ M=1553 MeV might possibly be identified with Roper resonance [30-34]. The structure of the Roper resonance can be described as the mixing of three-particle system and the nucleon pentaquark.

## IV Conclusion.

In strongly bound system of light quarks such as the baryons consideration, where $p/m \approx 1$ the approximation of nonrelativistic kinematics and dynamics not justified.

In our relativistic five-quark model (Faddeev – Yakubovsky type approach) we calculated the masses of low-lying pentaquarks. We used $SU(3)_f$ symmetry. The quark amplitudes obey the global color symmetry. The masses of the constituent quarks are equal to 405 MeV. We considered the scattering amplitudes of the constituent quarks. The poles of these amplitudes determine the masses of low-lying pentaquarks. The derived five-quark amplitude consists of three subamplitudes: $BM$, $Mqqq$, $Bq\bar{q}$, where $B$ and $M$ are the baryon and the meson respectively.

Unlike mesons, all half-integral spin and parity quantum numbers are allowed in the baryon sector, so that experiments search for such pentaquark are not simple. Furthermore, no decay channels are a priori forbidden. These two facts make identification of a pentaquark difficult.

We manage with the quarks as with real particles. However, in the soft region the quark diagrams should be treated as spectral integrals of the quark mass with the spectral density $\rho(m^2)$: the integration of the quark mass in the amplitudes eliminates the quark singularities and introduces the hadron ones. One can believe that the approximation:

$$\rho(m^2) \Rightarrow \delta(m^2 - m_q^2) \tag{20}$$

could be possible for the low-lying hadrons (here $m_q$ is the "mass" of the constituent quark). We hope that the approach given by (20) is sufficiently good for the calculation of the low-lying pentaquarks being carried out here.

Acknowledgments.

One of authors S.M. Gerasyuta is indebted to Institut fur Kernphysik Forschungzentrum Julich for the hospitality where this work has been beginning. The authors would like to thank T. Barnes, D.I. Diakonov, S. Krewald, N.N. Nikolaev, P.R. Page for useful discussions. This research was supported in part by Russian Ministry of Education, Program "Universities of Russia" under Contract № 01.20.00. 06448.

APPENDIX A

We can go over from integration with respect of the cosines of angles to integration with respect to the energy variables by using the relations:

$$s'_{13} = 2m^2 + \frac{s_{123} - s'_{12} - m^2}{2} + \frac{z_1}{2}\sqrt{\frac{s'_{12} - 4m^2}{s'_{12}}[(s_{123} - s'_{12} - m^2)^2 - 4s'_{12}m^2]} \tag{A1}$$

$$s'_{24} = 2m^2 + \frac{s_{124} - s'_{12} - m^2}{2} + \frac{z_1}{2}\sqrt{\frac{s'_{12} - 4m^2}{s'_{12}}[(s_{124} - s'_{12} - m^2)^2 - 4s'_{12}m^2]} \tag{A2}$$

$$z = \frac{2s'_{12}(s_{1234} + s'_{12} - s_{123} - s_{124}) - (s_{123} - s'_{12} - m^2)(s_{124} - s'_{12} - m^2)}{\sqrt{[(s_{123} - s'_{12} - m^2)^2 - 4m^2 s'_{12}][(s_{124} - s'_{12} - m^2)^2 - 4m^2 s'_{12}]}} \tag{A3}$$

$$s'_{134} = m^2 + s'_{34} + \frac{s_{1234} - s'_{12} - s'_{34}}{2} + \frac{z_3}{2}\sqrt{\frac{s'_{12} - 4m^2}{s'_{12}}\left[(s_{1234} - s'_{12} - s'_{34})^2 - 4s'_{12}s'_{34}\right]} \quad (A4)$$

$$s'_{13} = 2m^2 + \frac{s'_{134} - s'_{34} - m^2}{2} + \frac{z_4}{2}\sqrt{\frac{s'_{34} - 4m^2}{s'_{34}}\left[(s'_{134} - s'_{34} - m^2)^2 - 4m^2 s'_{34}\right]} \quad (A5)$$

The integration in consideration take on the physical region, where $-1 \leq z_i \leq 1$ ($i = 1, 2, 3, 4$). Then one can define the integration region on the invariant variables. Therefore for $s'_{124}$ we have condition $0 \leq z^2 \leq 1$,

$$s'^{\pm}_{124} = s'_{12} + m^2 + \frac{(s_{1234} - s_{123} - m^2)(s_{123} + s'_{12} - m^2)}{2s_{123}} \pm$$
$$\pm \frac{1}{2s_{123}}\sqrt{[(s_{123} - s'_{12} - m^2)^2 - 4m^2 s'_{12}][(s_{1234} - s_{123} - m^2)^2 - 4m^2 s_{123}]} \quad (A6)$$

and the region of integration on $s'_{12}$ in $J_3$:

$$\tilde{\Lambda} = \begin{cases} \Lambda, & \text{if } \Lambda \leq (\sqrt{s_{123}} + m)^2 \\ (\sqrt{s_{123}} + m)^2, & \text{if } \Lambda > (\sqrt{s_{123}} + m)^2 \end{cases} \quad (A7)$$

APPENDIX B

The integration contour 1 (Fig. 3) corresponds to the connection $s_{123} < (\sqrt{s_{12}} - m)^2$, the contour 2 is defined by the connection $(\sqrt{s_{12}} - m)^2 < s_{123} < (\sqrt{s_{12}} + m)^2$. The point $s_{123} = (\sqrt{s_{12}} - m)^2$ is not singular, that the round of this point at $s_{123} + i\varepsilon$ and $s_{123} - i\varepsilon$ gives identical result. $s_{123} = (\sqrt{s_{12}} + m)^2$ is the singular point, but in our case the integration contour can not pass through this point that the region in consideration is situated below the production threshold of the four particles $s_{1234} < 16m^2$. The similar situation for the integration over $s_{13}$ in the function $J_3$ is occurred. But the difference consists of the given integration region that is conducted between the complex conjugate points (contour 2 Fig. 3). In Fig. 3, 4b, 5 the dotted lines define the square-root cut of the Chew-Mandelstam functions. They correspond to two-particles threshold and also three-particles threshold in Fig. 4a. The integration contour 1 (Fig. 4a) is determined by $s_{1234} < (\sqrt{s_{12}} - \sqrt{s_{34}})^2$, the contour 2 corresponds to the case $(\sqrt{s_{12}} - \sqrt{s_{34}})^2 < s_{1234} < (\sqrt{s_{12}} + \sqrt{s_{34}})^2$. $s_{1234} = (\sqrt{s_{12}} - \sqrt{s_{34}})^2$ is not singular point, that the round of this point at $s_{1234} + i\varepsilon$ and $s_{1234} - i\varepsilon$ gives identical results. The integration contour 1 (Fig. 4b) is determined by region $s_{1234} < (\sqrt{s_{12}} - \sqrt{s_{34}})^2$ and $s_{134} < (\sqrt{s_{34}} - m)^2$, the integration contour 2 corresponds to $s_{1234} < (\sqrt{s_{12}} - \sqrt{s_{34}})^2$ and $(\sqrt{s_{34}} - m)^2 \leq s_{134} < (\sqrt{s_{34}} + m)^2$. The contour 3 is defined by $(\sqrt{s_{12}} - \sqrt{s_{34}})^2 < s_{1234} < (\sqrt{s_{12}} + \sqrt{s_{34}})^2$. Here the singular point would be $s_{134} = (\sqrt{s_{34}} + m)^2$. But in our case this point is not achievable, if one has the condition $s_{1234} < 16m^2$. We have to consider the integration over $s_{24}$ in the function $J_3$. While $s_{124} < s_{12} + 5m^2$ the integration is conducted along the complex axis (the contour 1, Fig. 5). If we come to the point $s_{124} = s_{12} + 5m^2$, that the output into the square-root cut of Chew-Mandelstam function (contour 2, Fig. 5) is occurred. In this case the part of the integration contour in nonphysical region is situated and the integration contour along the real axis is conducted. The other part of integration contour corresponds to physical regions. This part of integration contour along the complex axis is conducted. The suggested calculation shows that the contribution of the integration over the nonphysical region is small [28, 29].

APPENDIX C

We considered the algebraic equations and determinants, which allow one to calculate the poles of reduced amplitudes $\alpha_1$, $\alpha_2$, $\alpha_3$ for the low-lying pentaquark. If we use the diquark with $J^P = 0^+$ ($l, p, r$ are equal 1 or 3), we can calculate the spectrum of $N$ pentaquarks. If we use the diquark with $J^P = 1^+$ ($l, p, r$ are equal 2 or 3), we can calculate the spectrum of $\Delta$ - isobar pentaquarks.

Figure 1

$$\alpha_1 = \lambda_1 + 6\alpha_3 J_2(3,1,1) + 2\alpha_2 J_2(3,1,3) + 6\alpha_2 J_1(3,3,1) + 2\alpha_3 J_1(3,1,1)$$
$$\alpha_2 = \lambda_2 + 4\alpha_1 J_3(3,3,1) \quad \text{(C1)}$$
$$\alpha_3 = \lambda_3 + 2\alpha_1 J_3(1,1,3)$$
$$D(s) = 1 - 8J_3(3,3,1)\{J_2(3,1,3) + 3J_1(3,3,1)\} - 4J_3(1,1,3)\{3J_2(3,1,1) + J_1(3,1,1)\}$$

Figure 2

$$\alpha_1 = \lambda_1 + 2\alpha_2 J_2(3,1,3) + 6\alpha_2 J_1(3,3,1)$$
$$\alpha_2 = \lambda_2 + 4\alpha_1 J_3(3,3,1) \quad \text{(C2)}$$
$$D(s) = 1 - 8J_3(3,3,1)\{J_2(3,1,3) + 3J_1(3,3,1)\}$$

Functions $J_1(l,p,r)$, $J_2(l,p,r)$ and $J_3(l,p,r)$ correspond to (16) – (18), $l, p, r = 1, 2, 3$.

APPENDIX D

The vertex functions are shown in Table 3. The two-particle phase space for the equal quark masses is defined as:

$$\rho_n(s_{ik}, J^{PC}) = \left(\alpha(J^{PC},n)\frac{s_{ik}}{4m^2} + \beta(J^{PC},n)\right)\sqrt{\frac{s_{ik} - 4m^2}{s_{ik}}},$$

The coefficients $\alpha(J^{PC},n)$ and $\beta(J^{PC},n)$ are given in Table 4.

Table I. Low-lying nucleon pentaquark masses and contributions of subamplitudes $BM$, $Mqqq$ and $Bq\bar{q}$ to pentaquark amplitude in % (diquark with $J^P = 0^+$).

| Fig. № | Meson $J^{PC}$ | $J^P$ | Mass, MeV | $BM$ | $Mqqq$ | $Bq\bar{q}$ |
|---|---|---|---|---|---|---|
| 1 | $0^{++}$ | $\frac{1}{2}^+$ | 1553 $P_{11}(1440)$ | 49.35 | 38.54 | 12.11 |
| 1 | $1^{++}$ | $\frac{1}{2}^+, \frac{3}{2}^+$ | 1650 $P_{11}(1710)$ $P_{13}(1720)$ | 48.36 | 38.75 | 12.89 |
| 1 | $2^{++}$ | $\frac{3}{2}^+$ | 1875 $P_{13}(1900)$ | 45.17 | 39.06 | 15.77 |
| 2 | $2^{++}$ | $\frac{5}{2}^+$ | 1990 $F_{15}(1990)$ | 49.08 | 50.92 | - |
| 1 | $0^{-+}$ | $\frac{1}{2}^-$ | 1378 $S_{11}(1535)$ | 55.53 | 35.84 | 8.63 |
| 1 | $1^{--}$ | $\frac{1}{2}^-, \frac{3}{2}^-$ | 1814 $S_{11}(1650)$ $D_{13}(1700)$ | 44.87 | 39.52 | 15.61 |

Parameters of model: quark mass $m = 405$ MeV, cut-off parameter $\Lambda = 22$; gluon coupling constant $g = 0.357$. Experimental mass values of nucleon pentaquark are given in parentheses [19].

Table II. Low-lying $\Delta$ - isobar pentaquark masses and contributions of subamplitudes $BM$, $Mqqq$ and $Bq\bar{q}$ to pentaquark amplitude in % (diquark with $J^P = 1^+$).

| Fig. № | Meson $J^{PC}$ | $J^P$ | Mass, MeV | $BM$ | $Mqqq$ | $Bq\bar{q}$ |
|---|---|---|---|---|---|---|
| 1 | $0^{++}$ | $\frac{3}{2}^+$ | 1290 ( - ) | 75.05 | 16.62 | 8.33 |
| 1 | $1^{++}$ | $\frac{1}{2}^+, \frac{3}{2}^+, \frac{5}{2}^+$ | 1580 $P_{31}(1750)$ $P_{33}(1600)$ ( - ) | 58.78 | 31.24 | 9.98 |
| 1 | $2^{++}$ | $\frac{1}{2}^+, \frac{3}{2}^+, \frac{5}{2}^+$ | 1845 $P_{31}(1910)$ $P_{33}(1920)$ $F_{35}(1905)$ | 50.70 | 36.12 | 13.18 |
| 2 | $2^{++}$ | $\frac{7}{2}^+$ | 1970 $F_{37}(1950)$ | 52.26 | 47.74 | - |
| 1 | $0^{-+}$ | $\frac{3}{2}^-$ | 1150 ( - ) | 71.19 | 22.92 | 5.89 |
| 1 | $1^{--}$ | $\frac{1}{2}^-, \frac{3}{2}^-, \frac{5}{2}^-$ | 1782 $S_{31}(1620)$ $D_{33}(1700)$ $D_{35}(1930)$ | 51.51 | 35.76 | 12.73 |

Parameters of model: quark mass $m = 405$ MeV, cut-off parameter $\Lambda = 32.4$; gluon constant $g = 0.357$. Experimental mass values of $\Delta$ - isobar pentaquarks are given in parentheses [19].

Table III. Vertex functions

| $J^{PC}$ | $G_n^2$ |
|---|---|
| $0^+$ (n=1) | $4g/3 - 8gm^2/(3s_{ik})$ |
| $1^+$ (n=2) | $2g/3$ |
| $0^{-+}$ (n=3) | $8g/3 - 16gm^2/(3s_{ik})$ |
| $1^{--}$ (n=3) | $4g/3$ |
| $0^{++}$ (n=3) | $8g/3$ |
| $1^{++}$ (n=3) | $4g/3$ |
| $2^{++}$ (n=3) | $4g/3$ |

Table IV. Coefficient of Chew-Mandelstam functions for n = 3 (meson states) and diquarks n = 1 ($J^P = 0^+$), n = 2 ($J^P = 1^+$).

| $J^{PC}$ | n | $\alpha(J^{PC}, n)$ | $\beta(J^{PC}, n)$ |
|---|---|---|---|
| $0^{++}$ | 3 | 1/2 | -1/2 |
| $1^{++}$ | 3 | 1/2 | 0 |
| $2^{++}$ | 3 | 3/10 | 1/5 |
| $0^{-+}$ | 3 | 1/2 | 0 |
| $1^{--}$ | 3 | 1/3 | 1/6 |
| $0^+$ | 1 | 1/2 | 0 |
| $1^+$ | 2 | 1/3 | 1/6 |

Figure captions.

Fig.1. Graphic representation of the equations for the five-quark subamplitudes $A_1(s, s_{1234}, s_{12}, s_{34})$ ($BM$), $A_2(s, s_{1234}, s_{12}, s_{123})$ ($Mqqq$), and $A_3(s, s_{1234}, s_{25}, s_{125})$ ($Bq\bar{q}$) using the low-lying mesons with $J^{PC} = 0^{++}, 1^{++}, 2^{++}, 0^{-+}, 1^{--}$ and diquarks with $J^P = 0^+, 1^+$.

Fig.2. Graphic representation of the equations for the five-quark subamplitudes $A_1(s, s_{1234}, s_{12}, s_{34})$ ($BM$), $A_2(s, s_{1234}, s_{12}, s_{123})$ ($Mqqq$).

Fig. 3. Contours of integration 1, 2 in the complex plane $s_{13}$ for the functions $J_1$, $J_3$.

Fig. 4. Contours of integration 1, 2, 3 in the complex plane $s_{134}$ (a) and $s_{13}$ (b) for the function $J_2$.

Fig. 5. Contours of integration 1, 2 in the complex plane $s_{24}$ for the function $J_3$.

References.

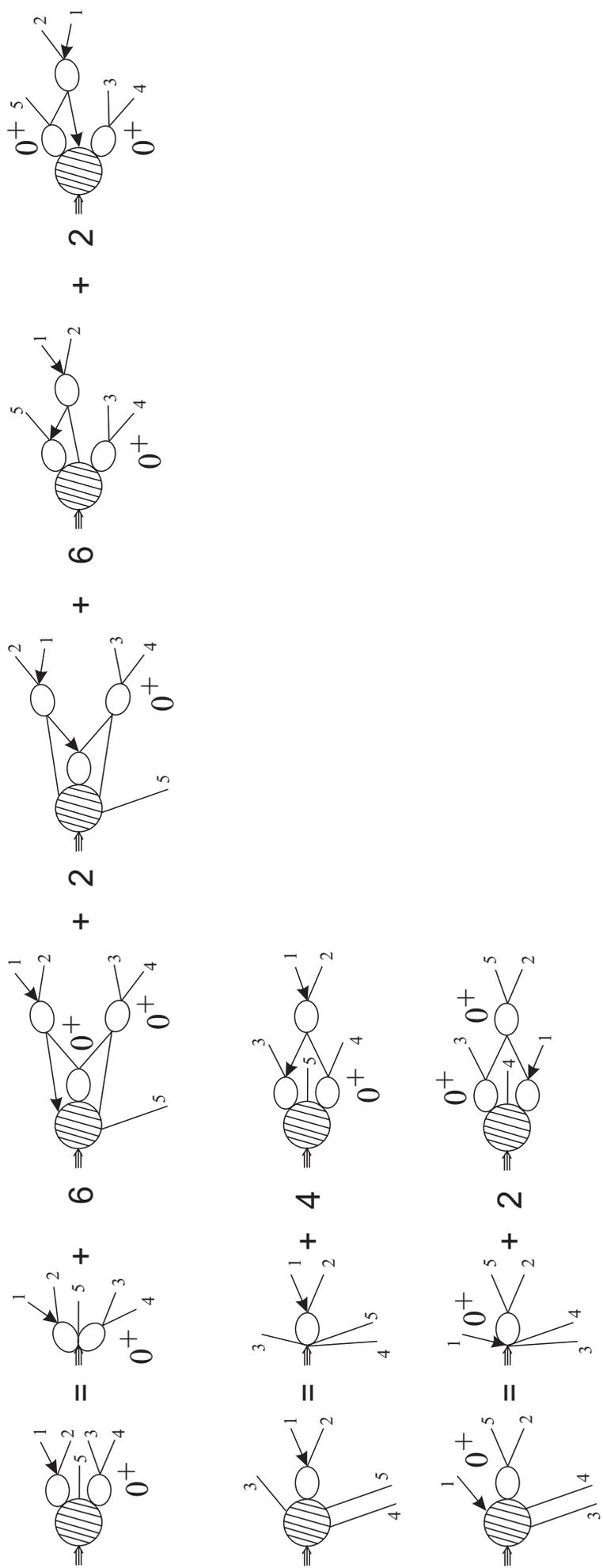

Fig.1

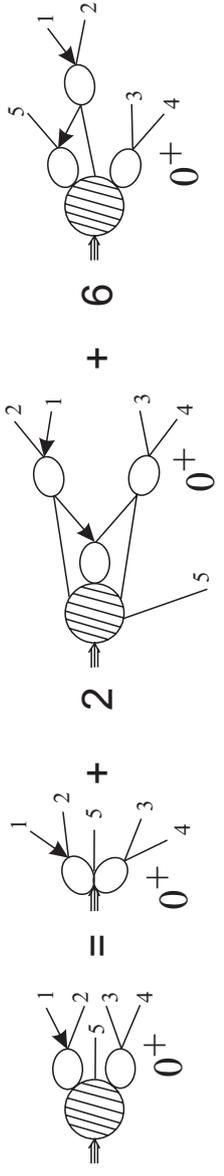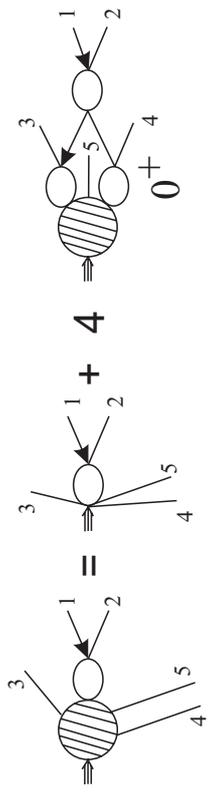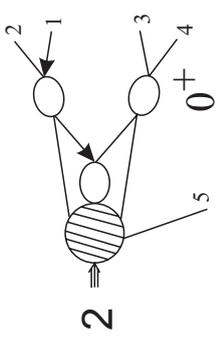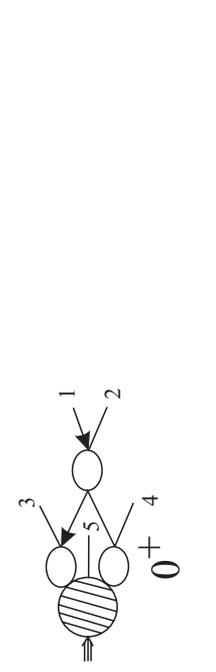

Fig.2

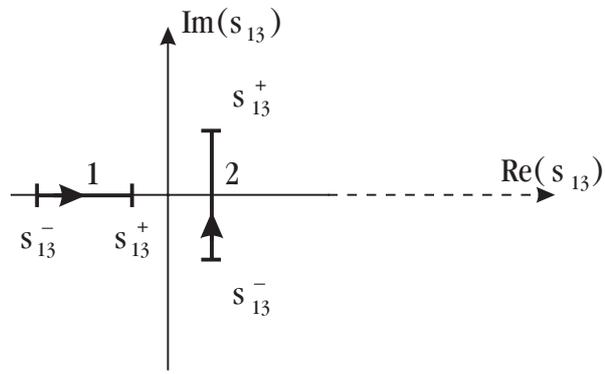

Fig. 3

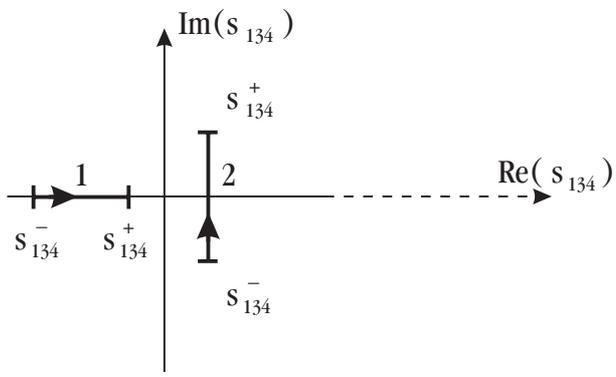

Fig. 4a

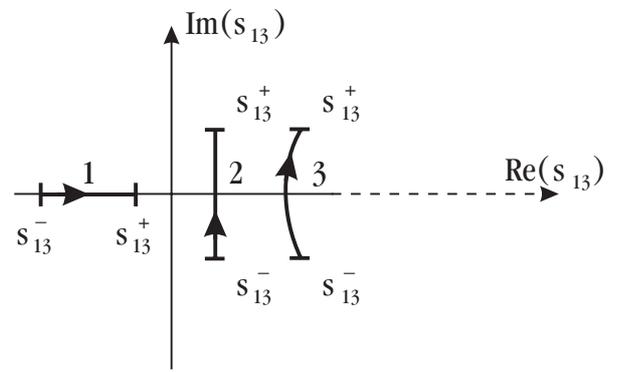

Fig. 4b

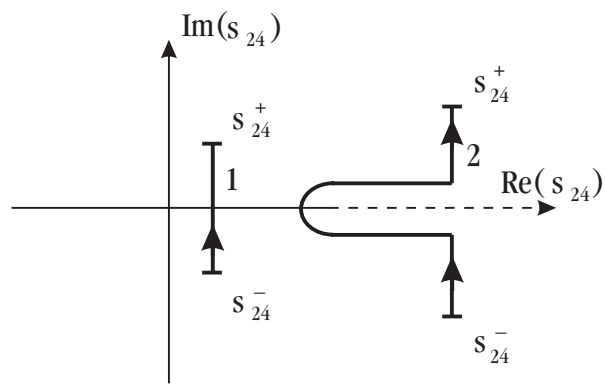

Fig. 5